\documentclass[aps,prd,twocolumn,superscriptaddress,nofootinbib]{revtex4-2}

\usepackage{graphicx}
\usepackage{amssymb,amsmath,amsfonts}
\usepackage{xcolor}
\usepackage[colorlinks=true,linktocpage=true,linkcolor=blue,citecolor=blue,urlcolor=blue]{hyperref}
\usepackage{slashed}

\def\tra{{\rm Tr}}
\def\traD{\tra_\sl[\mathrm{D}]}
\DeclareMathOperator\arctanh{arctanh}
\DeclareMathOperator\arccoth{arccoth}
\DeclareMathOperator\arcsinh{arcsinh}
\def\sm[#1]{{\scalebox{.7}{$\scriptscriptstyle #1$}}}
\def\sv[#1]{{\scalebox{.9}{$\scriptscriptstyle #1$}}}
\def\sl[#1]{{\scalebox{1.1}{$\scriptscriptstyle #1$}}}
\def\per{\!\sl[\perp]}
\makeatletter
\newcommand{\npar}{\mathrel{\mathpalette\new@parallel\relax}}
\newcommand{\new@parallel}[2]{%
  \begingroup
  \sbox\z@{$#1T$}
  \resizebox{!}{\ht\z@}{\raisebox{\depth}{$\m@th#1 ||$}}%
  \endgroup
}
\makeatother

\setcitestyle{citesep={,\kern-0.2em}} 
\begin{document}
\title{From Threshold Crossing to Wave-function Renormalization: Defining the Pion Mott Temperature in a Magnetic Field}
\author{Sidney S. Avancini} 
\affiliation{Departamento de F\'{\i}sica, Universidade Federal de Santa Catarina, 88040-900 Florian\'{o}polis, SC, Brazil}

\author{M\'aximo Coppola} \email{maximocoppola@cnea.gob.ar}
\affiliation{CNEA, Departamento de F\'{\i}sica Te\'orica (DFTIFSC), 1429 Buenos Aires, Argentina}

\author{Dyana C. Duarte} 
\affiliation{Departamento de F\'{i}sica, Universidade Federal de Santa Maria, 97105-900 Santa Maria, RS, Brazil}
\author{Ricardo L. S. Farias} 
\affiliation{Departamento de F\'{i}sica, Universidade Federal de Santa Maria, 97105-900 Santa Maria, RS, Brazil}

\author{Norberto Scoccola} 
\affiliation{CNEA, Departamento de F\'{\i}sica Te\'orica (DFTIFSC), 1429 Buenos Aires, Argentina}
\affiliation{CONICET, 1033 Buenos Aires, Argentina}
\author{William R. Tavares} 
\affiliation{CFisUC, Department of Physics, University of Coimbra, P-3004 - 516 Coimbra, Portugal}
%
%

\begin{abstract}
We investigate the dissociation of the neutral pion in hot
magnetized quark matter within the two-flavor Nambu--Jona-Lasinio
model. At zero magnetic field, the Mott temperature is
conventionally determined by $m_{\pi^0}(T_{\rm Mott})=2M(T_{\rm
Mott})$, above which a real pole ceases to exist. 
At finite magnetic field, Landau quantization replaces this single threshold by a hierarchy of quark--antiquark continua and generates multiple solutions of the pion pole equation, rendering a direct threshold-crossing criterion ambiguous since a real pion solution below the lowest nominal threshold always exists. 
We therefore propose to define the magnetic Mott temperature through the inflection point of the pion wave-function renormalization factor $Z_{\pi^0}(T,eB)$ of that lowest pole, corresponding to the fastest loss of its spectral function strength. 
The prescription reproduces the conventional Mott temperature as $eB\to 0$ and tracks the chiral pseudocritical temperature for both constant and magnetic-field-dependent couplings. 
Our results characterize pion dissociation in a magnetic field as a spectral crossover rather than a simple threshold crossing.
\end{abstract}
\maketitle
%
%
\noindent{\it Introduction} --- The dissociation of composite
states in a thermal medium is a central problem in strongly
interacting matter. In effective quark models of QCD, the pion
Mott transition is associated with the loss of the pion
bound-state character as chiral symmetry is
restored~\cite{Blaschke:2013zaa,Costa:2019bua}. At vanishing
magnetic field, the Mott temperature is conventionally defined
by~\cite{Quack:1994vc,Hufner:1996pq,Blaschke:2001yj,Costa:2002gk,Costa:2003uu,Hansen:2006ee}
\begin{equation}
m_\pi(T_{\rm Mott})=2M(T_{\rm Mott}) \, ,
\label{T_Mott_B0}
\end{equation}
where $M$ is the constituent-quark mass.
Below this threshold the pion is a bound state, whereas above it the mode enters the quark--antiquark continuum and becomes resonant~\cite{Hatsuda:1987kg,Asakawa:1989bq,Klevansky:1992qe,Zhuang:1994dw}.

A magnetic field qualitatively modifies this picture through Landau quantization.
In the presence of a magnetic field, the constituent quark spectrum is quantized into Landau levels (LLs)
\begin{equation}
E_{fn}^2 = p_z^2 + M_{f,n}^2 \ , \quad M_{f,n}^2 \equiv M_f^2+2n|q_f B| \, ,
\end{equation}
where $f$ stands for the flavor index. 
As a consequence, the quark-antiquark continuum is no longer characterized by a single threshold. 
Instead, an infinite sequence of thresholds emerges
\begin{equation}
m_{f,n}^{\rm (th)} = 2M_{f,n} \, ,
\end{equation}
corresponding to the opening of different decay channels.

This issue has been discussed in the context of different versions of the Nambu--Jona-Lasinio (NJL) model~\cite{Fayazbakhsh:2013cha,Zhang:2016qrl,Mao:2017wmq,Mao:2018dqe,Avancini:2018svs,Chaudhuri:2019lbw,Ghosh:2020qvg,Sheng:2020hge,YangShuYun:2021cfg,Mao:2019avr,Li:2021swv,Mei:2022dkd,Li:2023rsy,Mei:2026xlj,Tian:2026,Zhou:2026hqk,Li:2025wqb}.
As stated in Ref.~\cite{Mei:2026xlj}, the pole equation may
develop multiple branches due to this hierarchy of LL-thresholds.
In particular, a low mass stable solution associated to a real pole always exists for the neutral pion. 
This renders the identification of a Mott temperature from a simple crossing condition ambiguous. 
In fact, due to different judgment criteria, the reported results on the behavior of the magnetic Mott temperature vary even qualitatively. 
While at $B=0$ the pion polarization function at the single  threshold is continuous, at finite $B$ it diverges at the LL-thresholds due to the dimensional reduction induced by the magnetic field~\cite{Mao:2017wmq}. 
This implies that $m_{\pi^0}$ can never be equal to $2M_{f,n}$. 
In Refs.~\cite{Avancini:2018svs,Chaudhuri:2019lbw,Ghosh:2020qvg,Sheng:2020hge,YangShuYun:2021cfg} it was interpreted that a jump from the lowest stable solution to an excited state should occur when the mass of the lowest pion state gets numerically very close to the lowest threshold, defining $T_{\rm Mott}(B)$ as the temperature at which this jump occurs. 
It should be noted, however, that the fact that the low mass stable solution always exists makes the precise location of this jump ambiguous.
In any case, this definition leads to a magnetically catalyzed behavior of $T_{\rm Mott}(B)$, which closely follows the chiral pseudotransition temperature $T_{pc}(B)$, also catalyzed in the usual NJL model. 
Note that this resemblance is expected from the chiral limit, where $T_{\rm Mott}$ coincides with $T_{pc}$.
On the other hand, in Refs.~\cite{Mao:2017wmq,Mao:2018dqe,Mao:2019avr,Mei:2022dkd,Li:2023rsy}, the authors argue that, since the chiral condensate decreases with temperature, the mass of the stable pion mode should increase monotonically. 
Consequently, they consider any decreasing segment of this solution to be unphysical and define $T_{\rm Mott}$ as the temperature at which the real pole mass reaches its maximum value.
Unlike the results reported in Refs.~\cite{Avancini:2018svs,Sheng:2020hge,YangShuYun:2021cfg}, and in contrast to the behavior of $T_{pc}(B)$, this definition yields a decreasing $T_{\rm Mott}(B)$ with increasing magnetic field strength. 
Moreover, for sufficiently strong magnetic fields, it implies that the pion Mott transition occurs at temperatures deep within the chirally broken phase, where chiral symmetry remains substantially broken \cite{Mei:2026xlj,Li:2023rsy}.

In this work, we revisit the dissociation of the $\pi^0$ meson in hot magnetized quark matter within the two-flavor NJL model.
We show that the spectral weight~\cite{Hansen:2006ee,Mei:2026xlj,Zhou:2026hqk} carried by the lowest pion pole may become negligible even though a real subthreshold solution persists.
We therefore define $T_{\rm Mott}(B)$ through the inflection point of the pion wave-function renormalization factor $Z_{\pi^0}(T,B)$, which identifies the temperature of the fastest loss of the real pion pole strength.

This prescription always remains well defined at finite magnetic field and continuously reproduces the conventional Mott temperature in the limit $eB\to0$.
It also provides a natural description of pion dissociation as a spectral crossover driven by the redistribution of strength between the lowest pole and the quark--antiquark continuum, rather than as the crossing of a single kinematic threshold.

\noindent{\it Model Details} --- We employ the two-flavor NJL model~\cite{Buballa:2003qv,Klevansky:1992qe} in the presence of a uniform magnetic field. The Lagrangian density is given by
\begin{equation}
\mathcal{L} = \bar{\psi}
\left(i\slashed{D}-m_c \right) \psi +
G\left[ (\bar{\psi}\psi)^2 +
(\bar{\psi}i\gamma_5\vec{\tau}\psi)^2\right] \, ,
\end{equation}
where $m_c$ is the bare quark mass matrix in the isospin approximation,
$\psi=(\psi_u \quad \psi_d)^T$ is the quark field;  $G$ is the coupling constant and $\vec{\tau}$ are the Pauli matrices. Also, $D^\mu =(\partial^{\mu} + i \hat{Q} A^{\mu})$ is the covariant derivative, where $\hat{Q}$=diag($q_u,q_d$)=diag($2/3e,-1/3e$) is the quark charge matrix.
We will consider an homogeneous stationary magnetic field orientated along the 3-axis, $\vec{B}=B\hat{3}$.
Note that since we are dealing with neutral particles
all {possible Schwinger phases cancel and there is no need to specify a particular gauge for the calculations.
We adopt the magnetic field independent regularization~\cite{Avancini:2019wed} with 3D sharp-cutoff for the vacuum regularization.
For the model parameters we take: the cutoff, $\Lambda=664.3$ MeV,  the coupling constant, $G=2.06/\Lambda^2$, and the current quark masses, $m_c=5.0$ MeV~\cite{Buballa:2003qv}.

After bosonization and expansion around the mean-field solution, at mean-field level the constituent quark mass $M$ (which is equal for both flavors in the present model) is, as usual, determined self-consistently through the gap equation.
Meanwhile, the neutral-meson propagator, constructed in the spirit of the random phase approximation method~\cite{Vogl:1991qt,Klevansky:1992qe,Hatsuda:1994pi}, reads
\begin{equation}
\mathcal{G}(q) = \dfrac{2G}{1-2G\,\Pi(q)} \, ,
\end{equation}
where $\Pi(q)$ denotes the polarization function, which becomes complex above $2M$.
The corresponding explicit expressions for the gap equation and the polarization function are presented in the Appendixes.

The polarization function inherits the above mentioned structure and develops a hierarchy of branch points associated with the various Landau-level continua.
This feature is ultimately responsible for the appearance of multiple solutions in the pion pole equation reported in previous studies.

\begin{figure*}[ht]
\centering
\includegraphics[width=0.9\textwidth]{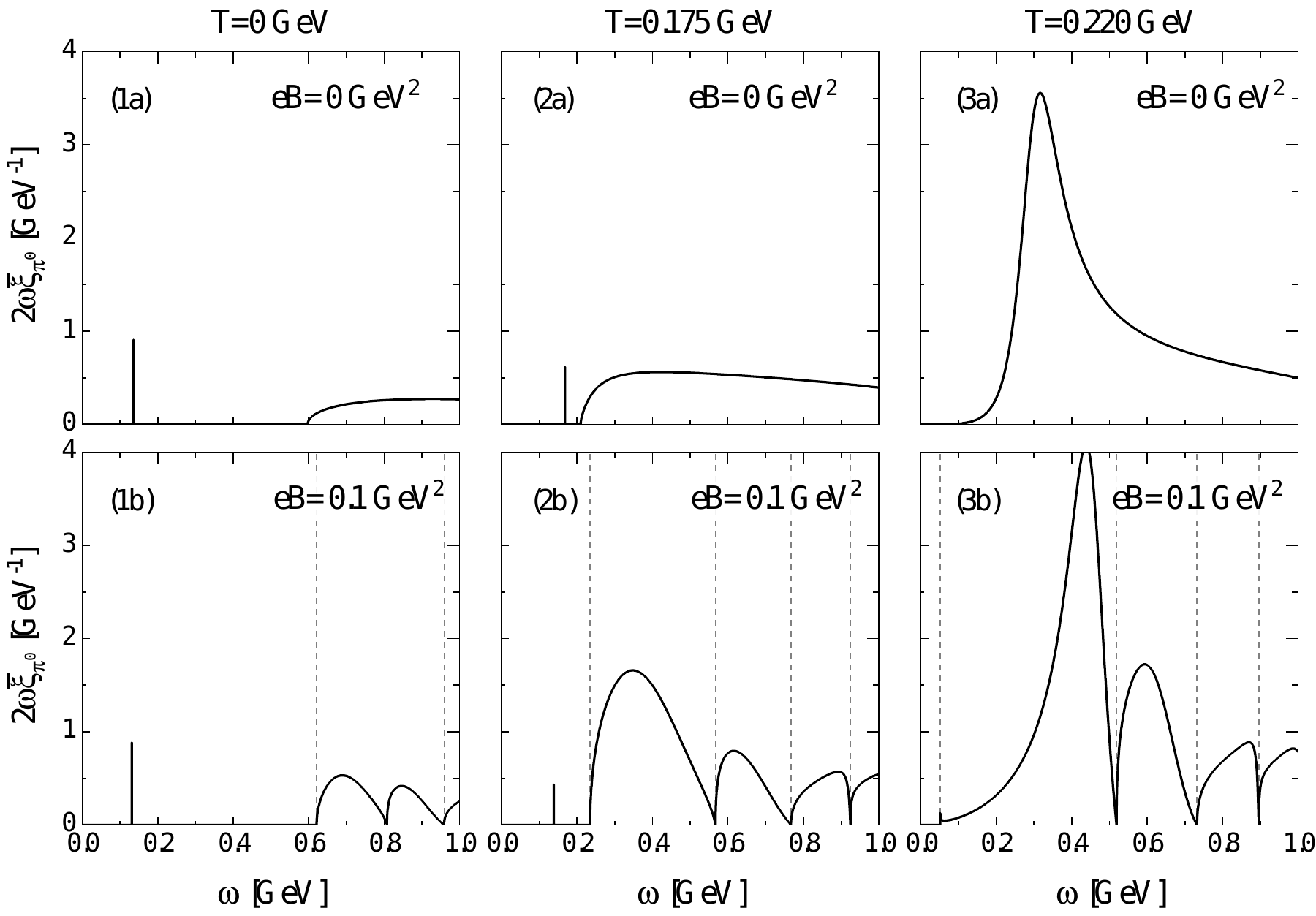}
\caption{$2\omega\bar{\xi}_{\pi^0}(\omega)$ as a function of $\omega$, where $\bar{\xi}_{\pi^0}$ is the renormalized spectral function of $\pi^0$.
The delta function at the real pole is represented by the value of $\bar{Z}_{\pi^0}$, not visible in subplot (3b) due to its negligible value.
The upper (lower) row correspond to $eB=0$~GeV$^2$ ($eB=0.1$~GeV$^2$), for temperatures $T=0$~GeV (first column), $T=0.175$~GeV (second column) and $T=0.220$~GeV (third column).
The vertical dashed lines correspond to the magnetic mass thresholds $2M_{d,n}$, with $n=0,1...$ the Landau level of constituent quarks.}
\label{fig:Spec2m_P0}
\end{figure*}

Spectral functions provide a powerful tool to study the distribution of many-particle states.
For mesons, the spectral function is related to the
imaginary part of the retarded propagator
\begin{equation}
\xi(q) \equiv \dfrac{1}{\pi} \, \mathrm{Im}\, \mathcal{G}(q^0+i\epsilon,\vec q\,)\, ,
\end{equation}
In what follows we will focus on rest-frame properties, described by setting $\vec{q}=\vec{0}$ and $q_0=\omega$.
If a bound state exists below $2M$, the spectral function will be given by the sum of this pole and the continuum
\begin{equation}
\xi(\omega) =  Z \, \delta(\omega^2-m_{\pi^0}^2)
+ \xi_{\rm c} (\omega^2) \Theta(\omega-2M)\, ,
\end{equation}
where $Z$ is the `wave function renormalization constant', defined by fixing the residue of the two-point function at the meson pole, while $\xi_c$ is the continuum contribution.

In a full renormalizable theory, the spectral density obeys a `sum rule': its integral is normalized to 1, as a consequence of the K\"all\'en–Lehmann representation~\cite{Peskin:1995ev}.
Since the NJL model is nonrenormalizable, a cutoff must be imposed, which determines the energy
range of applicability of the model.
If we define the upper bound of integration as $\Delta$, then, for fixed $B$ and $T$, the integral of $\xi$ will lead to a constant number $\mathcal{C}$ instead
\begin{equation}
\mathcal{C}(B,T) = \int_0^{\Delta^2} d\omega^2 \, \xi(\omega^2) \, .
\end{equation}
In order to compare results at different $B$ and $T$, we will normalize by $\mathcal{C}(B,T)$ each time to get renormalized quantities $\bar{x} = x/\mathcal{C}$, which fulfill
\begin{equation}
1 = \bar Z + \bar{A}[2M,\Delta] \ , \quad
\bar{A}[x,y] \equiv \int_{x}^y \, d\omega \, 2\omega\,
\bar \xi_{\rm c} (\omega^2) \, .
\label{prob}
\end{equation}
Since this relationship resembles that of a probability density function, we will interpret $\bar Z$ as the weight associated with the probability that the meson is in the lowest bound state.

\noindent{\it Results} --- In Fig.~\ref{fig:Spec2m_P0} we show the neutral pion integrand $2\omega \bar \xi_{\pi^0}(\omega^2)$ as a function of $\omega$ for different values of $B$ and $T$.
For visual comparison, the delta function at the real pole is represented by the value of its renormalized coefficient $\bar{Z}_{\pi^0}$.
While at $B=0$ a singular resonant state exists above the unique $2M$ threshold, at finite magnetic field the pion disassembles into a low bound state plus an infinite tower of excited states (resonances with finite decay widths).
Each excitation emerges between $2M_{d,n}$ and $2M_{d,n+1}$, and originates from the perpendicular momentum quantization of constituent quarks into Landau levels.
Note that, in the present model, thresholds of different flavors are degenerated as $m_{u,n}^{\rm (th)} = m_{d,2n}^{\rm (th)}$, since $|q_u| = 2 |q_d|$ and $M$ takes the same value for both flavors.

For a vanishing magnetic field, the bound state becomes a resonance at the Mott temperature~\cite{Bhattacharyya:1998ps,Wergieluk:2012gd,Blaschke:2016sqn}.
Remarkably, at $B\neq 0$ it can be analytically deduced from the polarization function that a bound solution always exists below $2M$.
The discontinuous jump reported in previous NJL calculations results from switching from this low-temperature branch to a different solution above the continuum threshold.
But since the bound solution always exists, no jump actually occurs: all states simply coexist as available static solutions to the pole equation.

Even though the lowest-lying stable solution persists in the magnetized medium, it constitutes a very weak bound state, since its mass lies very close to the $2M$ threshold.
This is because, as $T$ increases, $M$ decreases due to chiral symmetry restoration, lowering the $2M$ threshold and bringing it closer to the pion pole.
In fact, as $T$ increases the probability weight $\bar{Z}_{\pi^0}$ of the bound state is reduced against that of the continuum, which dominates at high-$T$.
For example, in Fig.~\hyperref[fig:Spec2m_P0]{1.(3b)} the peak of the bound state is not visible even though it is located at $\omega\simeq 0.0514~$GeV$<2M$,
because the height of $\bar{Z}_{\pi^0} \sim 10^{-4}$ is negligible.
Thus, at high temperatures the pion is more likely composed by a handful of excited states, each with different probability weight.

\begin{figure*}[htb]
\centering
\includegraphics[width=0.85\textwidth]{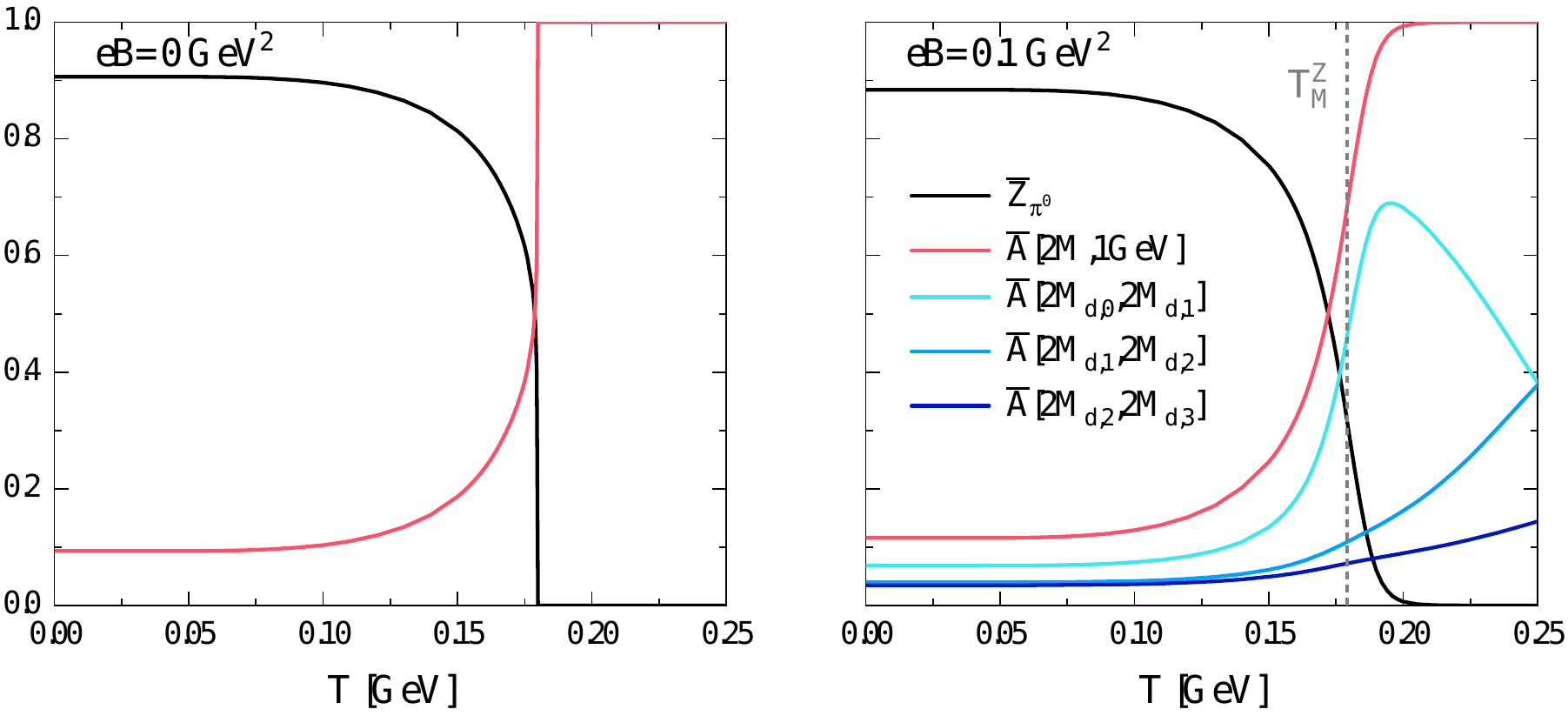}
\caption{Wave-function renormalization factor $Z_{\pi^0}$ (black line) and integrated spectral weights $A$, renormalized by their sum $\mathcal{C}(B,T)$, as functions of temperature.
The upper integration limit of $A$ is chosen as $\Delta=1$~GeV (red line).
Left panel: at $eB=0$, $\bar{Z}_{\pi^0}$ vanishes at the conventional Mott temperature.
Right panel: at finite magnetic field, exemplified by the value $eB=0.1$~GeV$^2$, $\bar{Z}_{\pi^0}$ decreases rapidly but remains nonzero.
The vertical dashed line indicates the Mott temperature obtained from its inflection point.
Blue lines indicate the probability weights of the first three resonant excitations.}
\label{fig:mott_definition}
\end{figure*}

To compare these weights, Fig.~\ref{fig:mott_definition} shows $\bar{Z}_{\pi^0}(T)$ and $\bar{A}(T)$, whose sum equals 1 according to Eq.~\eqref{prob},
taking $\Delta=1$~GeV as a reasonable upper energy limit of applicability of the model.
As $T$ increases, the bound-state weight $\bar{Z}_{\pi^0}$ decreases while the continuum contribution grows and dominates at high $T$.
The contributions of the three lowest excited states are also shown, revealing a progressive shift of spectral weight toward higher energies. Thus,
the pion identification becomes dynamical: at low $T$, it is dominated by the bound state, whereas at high $T$, its
spectral weight is increasingly associated with higher excited states.

At finite magnetic field, the existence of a nonvanishing bound solution for all values of $T$ naturally invalidates the conventional $B=0$ criterion of Eq.~\eqref{T_Mott_B0}, and raises the question of how to define the magnetic Mott transition.
For example, a sensible proposal consists of defining $T_{\rm Mott}$ as the point where the lines $\bar{Z}_{\pi^0}$ and $\bar{A}$ intersect.
There, the pion loses its distinct identity as a bound particle, since it is equally divided between a bound state and a continuum of quarks and antiquarks.
However, for effective descriptions such as the NJL model, this definition poses a problem due to the range of applicability of the model: excited states with energies above $\Delta$ are not taken into account.
Clearly, these are the ones that become more relevant with increasing $T$ and/or $B$.

To circumvent this issue, we will instead propose a definition based on the behavior of the real pole strength $\bar{Z}_{\pi^0}$.
As seen from the left panel of Fig.~\ref{fig:mott_definition}, at vanishing magnetic field $\bar{Z}$ decreases rapidly for increasing $T$ and vanishes at the Mott transition.
Thus, the Mott temperature can be equivalently defined as
\begin{equation}
Z_{\pi^0}(T_{\mathrm{Mott}},B=0)=0 \, .
\label{Z_Mott}
\end{equation}
At finite magnetic field, however, the situation changes qualitatively.
Since the lower bound state always exists, $\bar{Z}_{\pi^0}$ decreases strongly with temperature but does not vanish, as seen in the right panel of Fig.~\ref{fig:mott_definition}.
Therefore, the usual criterion of Eq.~\eqref{Z_Mott} cannot be applied.

We propose instead to define the magnetic Mott temperature through the inflection point of the wave-function renormalization factor
\begin{equation}
\left.
\dfrac{\partial^2 Z_{\pi^0}(T,B)}{\partial T^2}
\right|_{T=T_{\mathrm{Mott}}(B)}
=0 \, ,
\label{eq:mott_inflection}
\end{equation}
which corresponds to the temperature at which the absolute value of $\partial Z_{\pi^0}(T,B)/\partial T$ reaches its maximum.
This prescription identifies the temperature at which the real pole strength is lost most rapidly.
Below $T_{\mathrm{Mott}}(B)$, the pion mode retains a predominantly bound-state character, whereas above it the spectral weight is increasingly transferred to the quark--antiquark continuum.
The Mott temperature obtained from this criterion is shown in the right panel of Fig.~\ref{fig:mott_definition}.

The proposed definition has two main advantages.
First, in the $eB\rightarrow0$ limit it correctly converges to the $B=0$ value of $T_{\rm Mott}$, since the inflection point approaches the temperature at which $Z_{\pi^0}$ drops to zero.
At finite magnetic field, the Mott transition is therefore more naturally interpreted as a rapid spectral rearrangement rather than as the strict disappearance of the pion pole.
Second, it does not depend on the choice of a particular integration interval for the continuum spectral function.
In this sense, the inflection point provides a direct and model-independent operational criterion based on the thermal evolution of the pion pole itself.

Figure~\ref{fig:Tpc_TMott} shows the chiral pseudocritical temperature $T_{pc}$ and the Mott temperature $T_{\mathrm{Mott}}^{Z}$, defined from the inflection points of the chiral condensate and wave-function renormalization factor,
respectively, for both constant and magnetic-field-dependent couplings~\cite{Avancini:2016fgq,Coppola:2023mmq,Coppola:2025nus}
\begin{equation}
G(B)=G(0)\left[\kappa_1+(1-\kappa_1)e^{-\kappa_2(eB)^2}\right],
\end{equation}
with $\kappa_1=0.3206$ and $\kappa_2=1.31~\mathrm{GeV}^{-2}$, introduced to account for inverse magnetic catalysis~\cite{Bali:2011qj}.

\begin{figure}[ht]
\centering
\includegraphics[width=0.85\columnwidth]{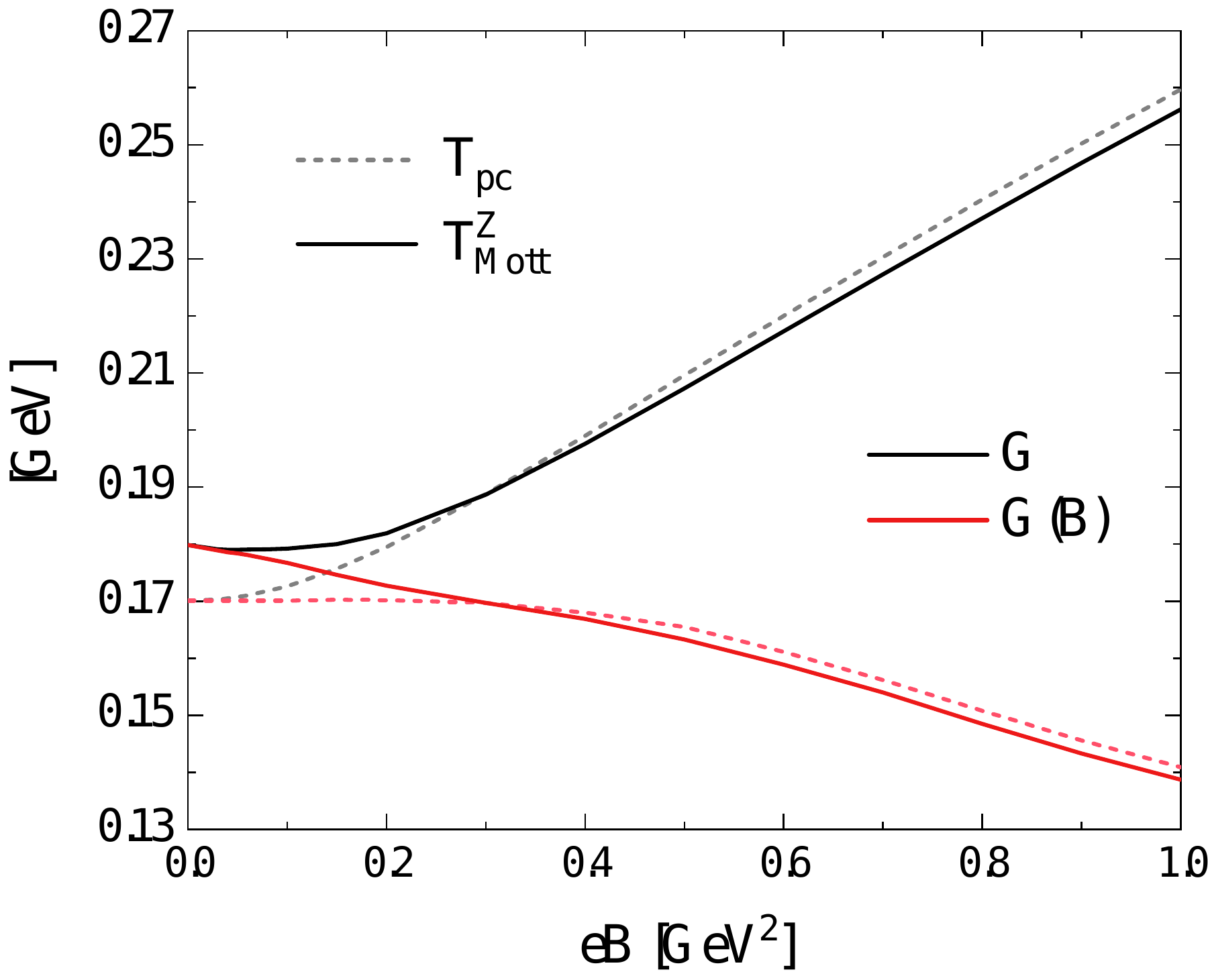}
\caption{
Chiral pseudocritical temperature, $T_{pc}$, and Mott temperature,
$T_{\mathrm{Mott}}^{Z}$, as functions of the magnetic field $eB$.
Dashed and solid lines denote $T_{pc}$ and $T_{\mathrm{Mott}}^{Z}$,
respectively. Black curves correspond to the constant coupling $G$,
whereas red curves correspond to the magnetic-field-dependent coupling
$G(B)$.
}
\label{fig:Tpc_TMott}
\end{figure}

For constant $G$, both temperatures increase with $eB$, reflecting the usual magnetic-catalysis behavior of the mean-field NJL model. 
When $G(B)$ is employed, both decrease, consistent with inverse magnetic catalysis.
In both cases, $T_{\mathrm{Mott}}^{Z}$ closely tracks $T_{pc}$, while remaining slightly lower at medium and large values of $eB$. 
Their proximity indicates a strong correlation between the rapid loss of pion pole strength and chiral symmetry restoration, whereas their small finite separation reflects the fact that they probe distinct spectral and thermodynamic properties. 
This behavior supports the use of the inflection point of $Z_{\pi^0}$ as an operational definition of the Mott temperature at finite magnetic field.

Before concluding it is interesting to discuss the temperature behavior of the sigma meson spectral function at finite $B$.
At zero magnetic field, the sigma meson degenerates with the pion above the pseudocritical chiral temperature due to the restoration of chiral symmetry.
Since the considered homogeneous magnetic field does not alter the nature of this transition, $\sigma$ and $\pi^0$ are also expected to be chiral partners within the magnetized medium.
As a consequence, the sigma should also undergo magnetic dissociation into an infinite tower of excited states.
This is shown in Fig.~\ref{fig:sigma} for $eB=0.1$~GeV$^2$, where we compare the sigma and pion spectral densities.
We can observe the same proliferation of excited states, each lying between two successive thresholds $2M_{d,n}$ and $2M_{d,n+1}$.
We also confirm the $\sigma-\pi^0$ degeneracy expected from the restoration of chiral symmetry: while at low temperatures ($T=0$ in the left panel) the spectral densities differ considerably,
at temperatures above $T_{\rm pc}$ ($T=0.22$~GeV in the right panel) both spectral densities coincide.
It is important to note that, contrary to the pion case, the sigma channel does not exhibit an analogous real pole below $2M$. This is due to absence
of a divergence at $\omega=2M$ in the corresponding polarization function. However, as it can be seen from Fig.~\ref{fig:sigma},
the participation of the low-energy pion root in the $\pi^0-\sigma$ degeneracy expected from chiral symmetry restoration is
completely negligible.

\begin{figure*}[t!]
\centering
\includegraphics[width=0.85\textwidth]{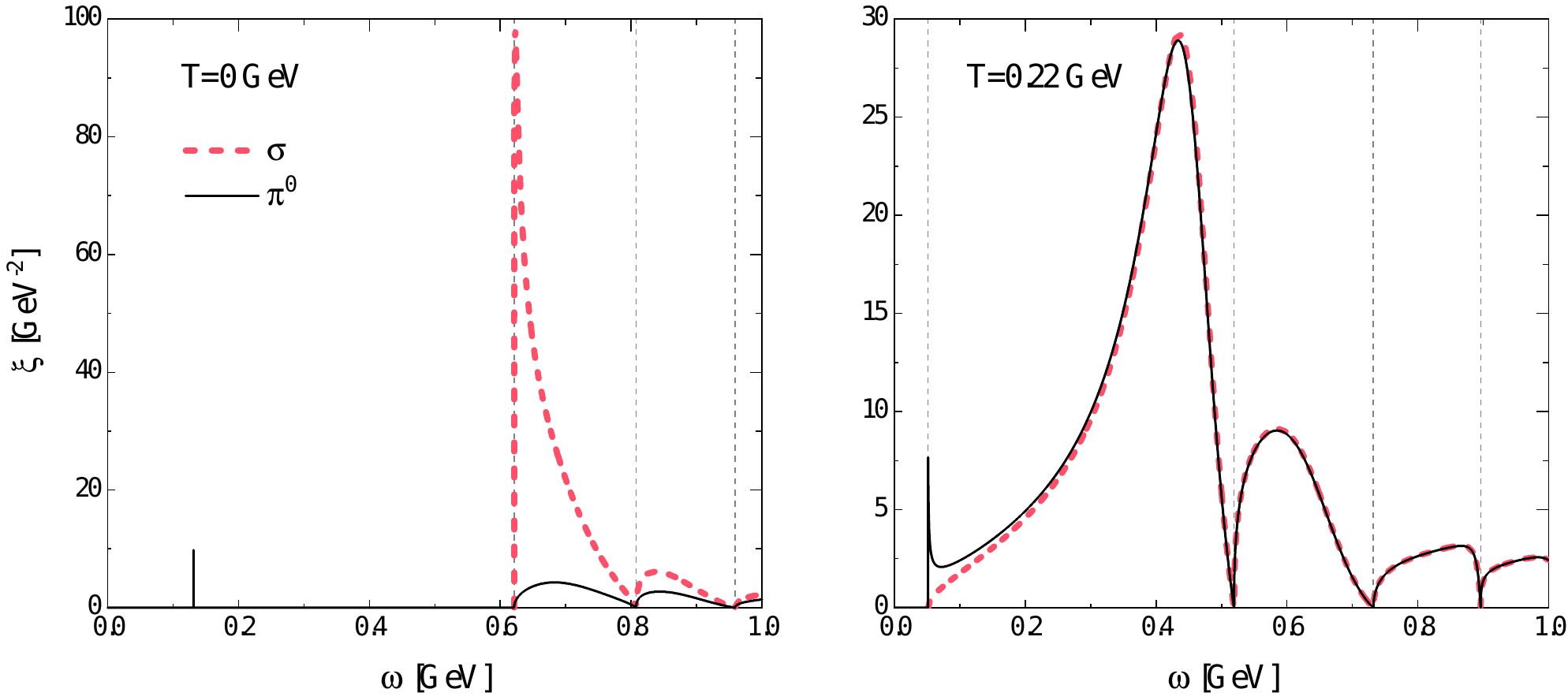}
\caption{
Spectral density of the sigma meson (red dashed lines) compared to the neutral pion one (full black lines) at $eB=0.1$~GeV$^2$ for two values of temperature: $T=0$ (left panel) and $T=0.22$~GeV (right panel).
The delta function at the real pole is represented by the value of $Z_{\pi^0}$, not visible in the right panel due to its negligible value.
The magnetic mass thresholds $2M_{d,n}$, corresponding to the quark Landau levels ($n=0,1,\ldots$), are depicted by vertical dashed lines.
}
\label{fig:sigma}
\end{figure*}

Finally, for completeness,  we have repeated our calculation using the Pauli-Villars regularization for the vacuum contribution
within the magnetic field independent regularization scheme \cite{Avancini:2019wed}.
The results so obtained are in qualitative agreement with those present above. This indicates that the findings in this letter
are basically independent of the vacuum regularization scheme.

\noindent {\it Conclusions} --- We have investigated the neutral pion Mott transition in hot magnetized quark matter within the two-flavor NJL model.
At vanishing magnetic field, the conventional condition
$m_{\pi^0}(T_{\rm Mott})=2M(T_{\rm Mott})$ coincides with the disappearance of the pion pole residue.
At finite magnetic field, however, Landau quantization generates multiple quark--antiquark thresholds and several solutions of the meson pole equations,
making a purely kinematic threshold-crossing criterion ambiguous.
In particular, the low bound-state solution always exists at finite magnetic field, but carries negligible spectral weight at high temperatures and has no analogous low-energy partner in the sigma channel.
By contrast, the higher pion and sigma branches approach degeneracy at high temperature, consistently with chiral symmetry restoration, although they belong to the quark--antiquark continuum.

Motivated by this spectral structure, we have proposed defining the magnetic Mott temperature, $T_{\rm Mott}^{Z}(B)$, through the inflection point of the pion wave-function renormalization factor $Z_{\pi^0}$ associated with the real pole, or equivalently through the maximum of $-dZ_{\pi^0}/dT$.
This criterion, which identifies the temperature at which the physical pion pole loses its spectral weight most rapidly, remains well defined when the lowest real solution persists, and continuously reproduces the conventional Mott temperature in the limit $B\to 0$.
Moreover, $T_{\rm Mott}^{Z}(B)$ closely follows the chiral pseudocritical temperature for both constant and magnetic-field-dependent couplings, reproducing magnetic catalysis and inverse magnetic catalysis, respectively.
These results support interpreting neutral pion dissociation in a magnetic background as a spectral crossover rather than as the crossing of a single quark--antiquark threshold.

This work was partially supported by Conselho Nacional de Desenvolvimento Cient\'ifico e Tecno\-l\'o\-gico  (CNPq), Grants No. 312032/2023-4, No. 402963/2024-5, 308963/2023-7 (S.S.A.) and 445182/2024-5 (R.L.S.F.), and  200037/2026-9 (W.R.T.); Funda\c{c}\~ao de Amparo \`a Pesquisa do Estado do Rio
Grande do Sul (FAPERGS), Grants No. 24/2551-0001285-0 (R.L.S.F.), No. 23/2551-0000791-6 and No. 23/2551-0001591-9 (D.C.D.). The work is also part of the project
Instituto Nacional de Ci\^encia e Tecnologia - F\'isica Nuclear e
Aplica\c{c}\~oes (INCT - FNA), Grants No. 464898/2014-5 and No. 408419/2024-5, and supported
by the Ser\-ra\-pi\-lhei\-ra Institute (Grant No. Serra -
2211-42230). R. L. S. F. acknowledges
the kind hospitality of the Center for Nuclear Research at Kent State University, where part of this work was
done. W.R.T. gratefully acknowledges the hospitality of the Centro de Física at the Universidade de Coimbra.

\appendix
\section{Mean field functions}

After bosonization, meson fields are expanded as fluctuations around mean field (MF) values $\bar \sigma $ and $\bar{\pi}_i=0$. 
At MF level, the action per unit volume is given by
\begin{equation}
\frac{S_{\mathrm{bos}}^{\mbox{\tiny MF}}}{V^{(4)}}\ =-\ \dfrac{\bar{\sigma}^2}{2G}-\dfrac{iN_{c}}{V^{(4)}}\sum_{f=u,d}\int d^{4}x\,d^{4}y
\ \traD\ln\left[\mathcal{S}_f(x,y)\right]^{-1}\, ,
\label{seff}
\end{equation}
where $\traD$ stands for the trace over Dirac space, and $\mathcal{S}_f(x,x')$
is the MF quark propagator in the presence of the magnetic field $\vec B=B\hat{3}$.
Its explicit expression can be written as
\begin{equation}
\mathcal{S}_f(x,y)=e^{i\Phi_f(x,y)}\int\!
\dfrac{d^{4}p}{(2\pi)^{4}}\ e^{-ip(x-y)}
\,\mathcal{S}_{f}(p)\, ,
\label{uno}
\end{equation}
The breaking of translational symmetry is manifested in the presence of the gauge-dependent Schwinger phase $\Phi_{f}(x,y)$.
The Landau level (LL) representation of the propagator is given by
\begin{align}
\mathcal{S}_{f}(p) = \, & 2 \, e^{-\vec p_{\per}^{\,2}/B_f} \, \sum_{n=0}^\infty \
\dfrac{(-1)^n}{p_{\npar}^2-2nB_f-M^2+i\epsilon} \, \times
\nonumber\\[2mm]
& \bigg\{ 
 (\slashed{p}_{\npar}+M) \left[ \Delta^s \, L_n^0\!\left( \dfrac{2\vec p_{\per}^{\,2}}{B_f} \right)-\Delta^{-s} \, L_{n-1}^0\!\left( \dfrac{2\vec p_{\per}^{\,2}}{B_f} \right) \right]
\nonumber\\[2mm]
& + 2 \, \vec{p}_{\per} \cdot \vec{\gamma}_{\per} \, L_{n-1}^1\!\left( \dfrac{2\vec p_{\per}^{\,2}}{B_f} \right) \bigg\} \, ,
\label{sfp_LL}
\end{align}
where $B_f = |B q_f|$ and $s={\rm sign}(B q_f)$, while $\Delta^s=(1+is\gamma^1\gamma^2)/2$. Also, $L^\alpha_n(x)$ are the generalized Laguerre polynomials, with $L_{-1}(x)=0$.
``Parallel'' and ``perpendicular'' four-vectors are defined as (similarly for $\gamma_{\npar}$ and $\gamma_{\per}$)
\begin{equation}
p_{\npar}^{\mu}=(p^{0},0,0,p^{3}) \quad , \quad
p_{\per}^{\mu}=(0,p^{1},p^{2},0)\, .
\end{equation}

The gap equation for the effective quark mass $M$, obtained by minimizing the free energy, is given by
\begin{equation}
M = m_c + 2GN_c M \sum_{f=u,d}  I_{1f} \, ,
\end{equation}
where the function $I_{1f}$ is proportional to the trace of the quark propagator
\begin{equation}
I_{1f} = \dfrac{1}{MV^{(4)}} \: \traD \int d^4x \, i \mathcal{S}_f(x,x) \, .
\end{equation}
After a rotation to Euclidean space ($p^0\to -ip_4$), the extension to finite temperature is performed through the replacements
\begin{align}
p_4 \ & \rightarrow \ \omega_\ell = (2 \ell +1)\pi T \, ,
\nonumber\\[2mm] 
\int_{-\infty}^\infty \dfrac{p_4}{2\pi} \ f(p_4) \ & \rightarrow \ T
\sum_{\ell=-\infty}^\infty \, f(\omega_\ell) \, .
\end{align}
When the sum over Matsubara frequencies is performed, $I_{1f}$ naturally separates as the sum of a thermomagnetic plus a purely magnetic term.
In the spirit of the adopted magnetic field independent regularization (MFIR) scheme, we subtract from the unregulated integral the $B\to 0$ limit and then add it in a regulated form using a sharp 3D-cutoff, leading to the decomposition 
\begin{align}
& \bullet \, I_{1f}^\sl[\rm B,T] =  I_{1f}^\sl[0,0] + I_{1f}^\sl[\rm mag,0] + I_{1f}^\sl[\rm mag,ther]
\nonumber\\[2mm]
& \bullet \, I_{1f}^\sl[0,0] = \dfrac{1}{2 \pi^2} \left[ \Lambda\, M_\Lambda + M^2 \ln\left( \dfrac{M}{\Lambda\, + M_\Lambda} \right) \right] 
\nonumber\\[2mm]
& \bullet \, I_{1f}^\sl[\rm mag,0] = \dfrac{B_f}{2\pi^2} \left[ \ln \Gamma(x_f) + \dfrac{1-2x_f}{2} \ln x_f + x_f - \dfrac{\ln{2\pi}}{2} \right] 
\nonumber\\[2mm]
& \bullet \, I_{1f}^\sl[\rm B,ther] =  - \dfrac{B_f}{2\pi^2} \displaystyle\sum_{n=0}^\infty g_n
\int_{-\infty}^\infty dp_3 \ \dfrac{n(E_{fn})}{E_{fn}} \, ,
\label{I1f}
\end{align}
where $M_\Lambda=\sqrt{\Lambda^2 + M^2}$ and $x_f=M^2/(2B_f)$.
Meanwhile $g_n=2-\delta_{n,0}$, $n(x)=(1+e^{x/T})^{-1}$ and also $E_{fn}^2=2nB_f+p_3^2+M^2$.

\section{Meson polarization function}

Meson masses are described from the second order expansion of the bosonized action.
In the neutral case, the contributions of Schwinger phases from each quark propagator cancel out since they correspond to the same quark flavor.
As a consequence, the polarization function is translational invariant, which leads to the conservation of momentum.
Taking the Fourier transform of the $M=\{\sigma,\pi^0\}$ fields to the momentum basis,
the corresponding transform of the quadratic action will be diagonal in momentum space
\begin{equation}
S_{M} \:=\: -\dfrac{1}{2} \int \dfrac{d^4q}{(2\pi)^4} \ \delta M(-q) \, \mathcal{G}^{-1}_M(q) \, \delta M(q)\: ,
\label{actionquadpi0p}
\end{equation}
where $\mathcal{G}_M(q)$ is the neutral meson propagator
\begin{equation}
\mathcal{G}_M(q) \:=\: 
\dfrac{2G}{1- 2 G\,\Pi_M(q)} \, .  
\label{pion_prop}
\end{equation}
After a rotation to Euclidean space ($q^0\to -iq_4$), the extension to finite temperature is performed through the replacements $q_4 \to \nu_m=2m\pi T$.
In order to calculate pole masses we take the meson at rest, $\vec q=\vec 0$, together with the analytic continuation $\nu_m \to im$.
Then, the polarization function can be written as
\begin{equation}
\Pi_M(m)= \sum_{f=u,d} \! N_c \left[ I_{1f}
-  (m^2-\varepsilon_M^2) \, I_{2f}(m) \right] \, ,
\end{equation}
where $\varepsilon_{\pi^0}=0$ while $\varepsilon_\sigma=2M$.
As in the MF level, to comply to the MFIR scheme we subtract from the unregulated integral the $B\to 0$ limit and then add it in a regulated form using a sharp 3D-cutoff.
As a result we obtain the decomposition 
\begin{equation}
I_{2f}^\sl[\rm B,T](m) = I_{2f}^\sl[0,0](m) +
I_{2f}^\sl[\rm mag,0](m) + I_{2f}^\sl[\rm B,ther](m) \, .
\end{equation}

\begin{widetext}
The vacuum $(B=T=0)$ terms read
\begin{align}
& \scalebox{0.6}{$\blacksquare$} \ \mathrm{Re} \left[ I_{2f}^\sl[0,0](m) \right] =
-\dfrac{1}{4\pi^2} \left[
\arcsinh \left(\dfrac{\Lambda}{M} \right) - F(m) \right]
\nonumber\\[4mm]
& \scalebox{0.6}{$\blacksquare$} \ F(m) =
\begin{cases}
  r_0\, \arctanh\left( \dfrac{1}{r_0} \, \dfrac{\Lambda}{M_\Lambda} \right)  
  & \!\! , \quad m > 2M_\Lambda
  \\[4mm]
  r_0\, \arccoth\left( \dfrac{1}{r_0} \, \dfrac{\Lambda}{M_\Lambda} \right)  
  & \!\! , \quad 2M <m < 2M_\Lambda
  \\[4mm]
  |r_0| \, \arctan\left( \dfrac{1}{|r_0|} \, \dfrac{\Lambda}{M_\Lambda} \right) 
  & \!\! , \quad  m < 2M
\end{cases}
\nonumber\\[4mm]
& \scalebox{0.6}{$\blacksquare$} \ \mathrm{Im} \left[ I_{2f}^\sl[0,0](m) \right] = -\dfrac{r_0}{8 \pi} \hspace{21.5mm} , \quad 2M <m < 2M_\Lambda 
\label{I2vacI}
\end{align}

Meanwhile, the magnetic contribution $(T=0)$ is given by
\begin{align}
& \scalebox{0.7}{$\clubsuit$} \ \mathrm{Re} \left[ I^\sl[\rm mag,0]_{2f}(m) \right] =
\dfrac{1}{8\pi^2} \, \times
\begin{cases}
\dfrac{1}{2} \, \displaystyle\int_{-1}^{1} dx \left[ \psi(\bar x_f) - \ln(\bar x_f) + \dfrac{1}{2 \bar x_f} \right]
\ , \quad m < 2M
\\[4mm]
\begin{aligned}
  & \dfrac{1}{2} \int_{-1}^{1} dx  \ \psi(\bar x_f + N + 1) - \ln(x_f) + 2 
 \nonumber \\[2mm]
 &  - r_0 \arccoth \left(\dfrac{1+r_0^2}{2\, r_0}\right) 
  + \dfrac{2 B_f }{m^2 }\sum_{n=0}^N \dfrac{ g_n }{r_n} \arccoth \left( \dfrac{1+r_n^2}{2 \, r_n} \right) \ , \quad  m > 2M
\end{aligned}
\end{cases} \\[2mm]
& \scalebox{0.7}{$\clubsuit$} \ \mathrm{Im} \left[ I^\sl[\rm mag,0]_{2f}(m) \right] = \dfrac{1}{8\pi} \left(r_0 -\dfrac{2 B_f }{m^2 } \sum_{n=0}^N \dfrac{g_n}{r_n} \right)  \ , \quad m > 2M
\end{align}
An important point is that the first term of the imaginary part of $I_{2f}^\sl[\rm mag,0](m)$ exactly cancels the imaginary vacuum contribution from $I_{2f}^\sl[\rm 0,0](m)$ in Eq.~\eqref{I2vacI}, as expected from the MFIR regularization.

Finally, the thermomagnetic contribution is given by
\begin{align}
& \scalebox{0.7}{$\spadesuit$} \ \mathrm{Re} \left[ I^\sl[\rm B,ther]_{2f} \right] =
\dfrac{B_f}{2\pi^2}  \times
\begin{cases}
\, \displaystyle \sum_{n=0}^\infty \, g_n \, \int_{-\infty}^\infty dp_3 \, \dfrac{1}{E_{fn}} \: \dfrac{n(E_{fn})}{4E_{fn}^2-m^2} \hspace{10mm} ,
\quad  m < 2M \\[8mm]
\, \displaystyle \sum_{n=N+1}^\infty \, g_n \, \int_{-\infty}^\infty dp_3 \, \dfrac{1}{E_{fn}} \: \dfrac{n(E_{fn})}{4E_{fn}^2-m^2} \, \\[6mm]
\, + \displaystyle \sum_{n=0}^N \, g_n \, \mathrm{PV}  
 \displaystyle\int_{-\infty}^\infty dp_3 \, \dfrac{1}{E_{fn}} \: \dfrac{n(E_{fn})}{4E_{fn}^2-m^2} \:\ , \quad m > 2M
\end{cases} \label{I2fBthR} \\[2mm]
& \scalebox{0.7}{$\spadesuit$} \ \mathrm{Im} \left[ I^\sl[\rm B,ther]_{2f} \right] = \dfrac{B_f}{2\pi m^2}
  \ n\left( \dfrac{m}{2} \right)\: \sum_{n=0}^N \, \dfrac{g_n}{r_n}
  \ , \quad m > 2M  \label{I2fBthI}
\end{align}
where PV denotes the Cauchy principal value of the integral.
For these expressions we have defined the following additional shorthand notation
\begin{align}
\bullet \ r_{n}^2=1-\dfrac{4(M^2+2nB_f)}{m^2} \:\ , \qquad
\bullet \ \bar x_f = \dfrac{1}{2B_f} \left(M^2-\dfrac{1-x^2}{4} \, m^2 \right) 
\ , \qquad 
\bullet \ N=\mathrm{Floor} \left[ \dfrac{m^2-4M^2}{8 B_f} \right]  \, .
\end{align}
\end{widetext}

%
\bibliography{cs2mss-references}
\end{document}